\documentstyle[twocolumn,seceq]{jpsj}

\input epsf

\def\ORDER#1{${\cal O}\left( #1 \right)$}

\def\KET#1{\vert{#1}\rangle}

\title{Quantum Spin Dynamics and Quantum Computation}

\def\runtitle{Quantum Spin Dynamics and Quantum Computation}
\def\runauthor{H. {\sc De Raedt} et al.}

\author{H. {\sc De Raedt}$^{1,}$\footnote{E-mail: deraedt@phys.rug.nl},
        A.H. {\sc Hams}$^1$,
        K. {\sc Michielsen}$^2$,
        S. {\sc Miyashita}$^3$ and
        K. {\sc Saito}$^3$
}

\inst{$^{1}$Institute for Theoretical Physics and Materials Science Centre \\
      $^{2}$Laboratory for Biophysical Chemistry \\
      University of Groningen, Nijenborgh 4 \\
      NL-9747 AG Groningen, The Netherlands \\
\vskip 0.5cm
      $^{3}$Department of Applied Physics, School of Engineering \\
      University of Tokyo, Bunkyo-ku, Tokyo 113  }

\recdate{
\today
}

\abst{
We describe a simulation method for a quantum spin model
of a generic, general purpose quantum computer.
The use of this quantum computer simulator is illustrated
through several implementations of Grover's database search algorithm.
Some preliminary results on the stability of quantum algorithms
are presented.
}

\kword
{
quantum computers, spin systems
}

\begin{document}
\sloppy
\maketitle

\section{Introduction}
The idea that a Quantum Computer (QC)
might be more powerful than an ordinary computer is based
on the notion that a quantum system can be in any superposition of states
and that interference of these states allows exponentially
many computations to be done in parallel.\cite{AHARONOVone}
A QC may solve certain computationally hard problems such as factoring integers
and searching databases faster than a conventional
computer.\cite{SHORone,CHUANGone,KITAEV,GROVERzero,GROVERone,GROVERtwo}
This intrinsic parallelism might be used to solve other
difficult problems as well, such as for example
the calculation of the physical properties of quantum many-body
systems.\cite{CERFone,ZALKAone,TERHALone,QCzero}
In fact, part of Feynman's original motivation
to consider QC's was that they might be used
as a vehicle to perform exact simulations of quantum mechanical
phenomena.\cite{FEYNMAN}

Theoretical work on quantum computation usually assumes the existence
of units that perform highly idealized unitary operations.
However, in practice these operations are difficult to realize.
Disregarding decoherence,
a hardware implementation of a QC will perform
unitary operations that are more complicated than those considered
in most theoretical work: In a QC the internal quantum dynamics
of each elementary constituent is a key ingredient of the QC itself.

This paper describes a simulator for a generic physical model
of a QC, strictly working according to the laws of quantum mechanics.
We implement Grover's database search
quantum algorithm (QA)\cite{GROVERone,GROVERtwo} using ideal and more realistic units,
such as those used in the 2-qubit
NMR QC.\cite{JONESone,JONEStwo,JONESthree,CHUANGtwo,CHUANGthree}

\section{Quantum Spin Dynamics}
Generically, hardware QC's are modeled
in terms of S=1/2 spins (qubits) that
evolve in time according to the time-dependent Schr\"odinger
equation (TDSE)
\begin{equation}\label{eq:tdse}
i{\partial \over\partial t} \KET{\Phi(t)}= H(t) \KET{\Phi(t)}
,
\end{equation}
in units such that $\hbar=1$ and where $\KET{\Phi(t)}$
describes the state of the whole QC at time $t$.
The time-dependent Hamiltonian $H(t)$ takes the form

\begin{eqnarray} \label{eq:fullham}
H(t)&=&-\sum_{j,k=1}^L\sum_{\alpha=x,y,z} J_{j,k,\alpha}(t) S_j^\alpha S_k^\alpha
\nonumber \\
&& -\sum_{j=1}^L\sum_{\alpha=x,y,z}
\left( h_{j,\alpha,0}(t)
\right. \nonumber \\
&& \left.
  + h_{j,\alpha,1}(t) \sin (f_{j,\alpha} t + \varphi_{j,\alpha})
\right) S_j^\alpha
,
\end{eqnarray}

\noindent
where the first sum runs over all pairs $P$ of spins,
$S_j^\alpha$ denotes the $\alpha$-th component of the spin-1/2
operator representing the $j$-th qubit,
$J_{j,k,\alpha}(t) $ determines the strength of the interaction between
the qubits labeled $j$ and $k$,
$h_{j,\alpha,0}(t)$ and $h_{j,\alpha,1}(t)$ are
the static (magnetic) and periodic (RF) field acting on the $j$-th spin respectively.
The frequency and phase of the periodic field are denoted by $f_{j,\alpha}$
and $\varphi_{j,\alpha}$.
The number of qubits is $L$ and the dimension of the Hilbert space $D=2^L$.
Hamiltonian~(\ref{eq:fullham}) captures the physics
of most candidate technologies for building QC's.

A QA for the QC modeled by~(\ref{eq:tdse}) and ~(\ref{eq:fullham})
consists of a sequence of elementary operations (EO's).
The action of an EO on the state $\KET{\Psi}$
of the quantum processor is determined by
the values of all the $J$'s and $h$'s (which are kept constant
during the operation) and the time interval it is active.
The input state $\KET{\Psi(t)}$ is transformed
into the output state $\KET{\Psi(t+\tau)}$ where $\tau$ denotes the
time it takes to complete the EO.
During this time interval the only time-dependence
of $H(t)$ is through the (sinusoidal) modulation of the fields on the spins.
The time evolution of the QC itself is governed by the TDSE~(\ref{eq:tdse}).

Formally the solution of~(\ref{eq:tdse}) can be expressed in terms of
the unitary transformation
$U(t+\tau,t)\equiv\exp_{+}(-i\int_{t}^{t+\tau} H(u) du)$,
where $\exp_{+}$ denotes the time-ordered exponential function.
Using the semi-group property of $U(t+\tau,t)$ we can write
\begin{eqnarray}
U(t+\tau,t)&=&U(t+m\delta,t+(m-1)\delta)
\cdots
\nonumber \\
&& U(t+2\delta,t+\delta)U(t+\delta,t),
\end{eqnarray}
where $\tau=m\delta$ ($m\ge1$). The standard procedure to construct
algorithms for solving the TDSE~(\ref{eq:tdse}) is to replace each $U(t+(n+1)\delta,t+n\delta)$
by a symmetrized product-formula
approximation\cite{HDRCPR,HUYGH,SUZUKI}.
For the case at hand a convenient choice is
(other decompositions\cite{PEDROone,SUZUKItwo}
work equally well but are somewhat less efficient for our purposes):
\begin{equation}
{U(t+(n+1)\delta,t+n\delta)}\approx
{\widetilde U(t+(n+1)\delta,t+n\delta)},
\end{equation}
\begin{eqnarray}
\lefteqn{{\widetilde U(t+(n+1)\delta,t+n\delta)}= } \nonumber \\
& &\quad\quad e^{-i\delta H_z(t+(n+1/2)\delta)/2}
e^{-i\delta H_y(t+(n+1/2)\delta)/2} \nonumber \\
& &\quad\quad \times e^{-i\delta H_x(t+(n+1/2)\delta)}
e^{-i\delta H_y(t+(n+1/2)\delta)/2} \nonumber \\
& &\quad\quad \times e^{-i\delta H_z(t+(n+1/2)\delta)/2},
\end{eqnarray}
where
\begin{eqnarray} \label{eq:hamil}
\lefteqn{H_\alpha(t) = } \nonumber \\
&&-\sum_{j,k=1}^L J_{i,j,\alpha} S_j^\alpha S_k^\alpha \nonumber \\
&&-\sum_{j=1}^L
\left( h_{j,\alpha,0} + h_{j,\alpha,1} \sin( f_{j,\alpha} t + \varphi_{j,\alpha})
\right) S_j^\alpha,
\end{eqnarray}
with $\alpha=x,y,z$. In~(\ref{eq:hamil}) the time dependence of the
$J$'s and the $h$'s has been omitted because these parameters are constant
during the execution of an EO.

Evidently ${\widetilde U(t+\tau,t)}$ is unitary by construction,
implying that the algorithm to solve the TDSE is unconditionally stable.\cite{HDRCPR}
It is easily shown that
 the algorithm is correct to second order in the time-step $\delta$.\cite{HDRCPR}
Furthermore ${\widetilde U(t+\tau,t)}$
can be used as a building block to construct higher-order
algorithms.\cite{HDRone,SUZUKItwoprime,HDRKRMone,SUZUKIthree}
In practice it is easy to find reasonable, relatively small, values of $m$ such that
the results obtained no longer depend on $m$ (and $\delta$).
Then, for all practical purposes, these results
are indistinguishable from the exact solution of the TDSE~(\ref{eq:tdse}).

It is customary to take as basis states $\{\KET{\phi_n}\}$
the direct product of the eigenvectors of the
$S_j^z$ (i.e. spin-up $\KET{\uparrow}_j$ and spin-down $\KET{\downarrow}_j$).
In this basis $e^{-i\delta H_z(t+(n+1/2)\delta)/2}$ changes the input state by altering
the phase of each of the basis vectors.
As $H_z$ is a sum of pair interactions it is trivial to rewrite this operation
as a direct product of 4x4 diagonal matrices (equivalent to the so-called
interaction-controlled phase shifts) and 4x4 unit matrices.
Hence the computation of $\exp(-i\delta H_z(t+(n+1/2)\delta)/2)\KET{\Psi}$ has been
reduced to the multiplication of two vectors, element-by-element.
The unitary matrix $e^{-i\delta H_y(t+(n+1/2)\delta)/2}$
can be written in a similar manner but the matrices that contain the
interaction-controlled phase-shift have to be replaced by
non-diagonal matrices. Although this does not present a real problem it is
more efficient and systematic to proceed as follows.
Let us denote by ${\cal X}$ (${\cal Y}$) the rotation by $\pi/2$ of all spins
about the $x$($y$)-axis. As
\begin{eqnarray} \label{eq:ehy}
e^{-i\delta H_y(t+(n+1/2)\delta)/2}&=&
{\cal X}{\cal X}^\dagger e^{-i\delta H_y(t+(n+1/2)\delta)/2}{\cal X}
{\cal X}^\dagger \nonumber \\
&=&{\cal X} e^{-i\delta H_z^\prime(t+(n+1/2)\delta)/2}{\cal X}^\dagger
,
\end{eqnarray}
it is clear that the action of
$e^{-i\delta H_y(t+(n+1/2)\delta)/2}$ can be computed by
applying to each qubit, the inverse of ${\cal X}$
followed by an interaction-controlled phase-shift and ${\cal X}$.
The prime in~(\ref{eq:ehy}) indicates that
$J_{i,j,z}$, $h_{i,z,0}$, $h_{i,z,1}$ and $f_{i,z}$
in $H_z(t+(n+1/2)\delta)$
have to be replaced by
$J_{i,j,y}$, $h_{i,y,0}$, $h_{i,y,1}$ and $f_{i,y}$ respectively.
A similar procedure is used to compute the action of
$e^{-i\delta H_x(t+(n+1/2)\delta)}$:
We only have to replace ${\cal X}$ by ${\cal Y}$.

By construction our algorithm to solve the TDSE~(\ref{eq:tdse}) for spin
model~(\ref{eq:fullham})
is a QA itself.
As a real QC operates on all qubits simultaneously
the operation counts for
$e^{-i\delta H_x(t+(n+1/2)\delta)}$, $e^{-i\delta H_y(t+(n+1/2)\delta)/2}$,
and $e^{-i\delta H_z(t+(n+1/2)\delta)/2}$
are \ORDER{(P+2)}, \ORDER{(P+2)}, and \ORDER{P}.

\section{Quantum Computation}
Using the QA outlined above quantum spin systems containing up to 24
S=1/2 spins can easily be simulated on present-day supercomputers.\cite{PEDROone}
Here our aim is to use this QA to simulate a recent realization
of a 2-qubit NMR QC\cite{JONESone,JONEStwo,CHUANGtwo,CHUANGthree}
and to execute QA's on this simulator.
In our calculations we will take the model parameters corresponding to
the NMR experiments of Refs.16,17 
in which the two nuclear spins
of the $^1$H and $^{13}$C atoms in a carbon-13 labeled chloroform
represent the two qubits.\cite{CHUANGtwo,CHUANGthree}
In the NMR set-up the molecules are placed in a strong static magnetic field in the
$+z$ direction. In the absence of interactions with other
degrees of freedom this spin-1/2 system can be modeled by the hamiltonian
\begin{equation} \label{eq:ham}
H =- J_{1,2,z} S_1^z S_2^z
- h_{1,z,0} S_1^z
- h_{2,z,0} S_2^z
,
\end{equation}
where $h_{1,z,0}/2\pi\approx 500 \hbox{MHz}$,
$h_{2,z,0}/2\pi\approx125 \hbox{MHz}$, and
$J_{1,2,z}/2\pi\approx-215 \hbox{Hz}$.\cite{CHUANGtwo}
As the antiferromagnetic interaction between the spins is much weaker
than the coupling to the external field and~(\ref{eq:ham})
is a diagonal matrix with respect to the basis states chosen, the ground state
of~(\ref{eq:ham}) is the state with the two spins up ($\KET{\uparrow\uparrow}$).
We denote this state
by $\KET{00}=\KET{\uparrow}\otimes\KET{\uparrow}=\KET{\uparrow\uparrow}$,
i.e. the state with spin up corresponds to a qubit $\KET{\uparrow}$.
A state of the $N$-qubit QC will be denoted by
$\KET{x_1x_2\ldots x_N}=\KET{x_1}\otimes\KET{x_2}\ldots\KET{x_N}$.

As usual it is expedient to write the TDSE for this NMR problem
in frames of reference rotating with the nuclear spin.
Substituting
$\KET{\Phi(t)}=e^{it(h_{1,z,0} S^z_1+h_{2,z,0} S^z_2)}\KET{\Psi(t)}$
the time evolution of $\Psi(t)$ in the absence of RF-fields
is governed by the hamiltonian $H =- J_{1,2,z} S_1^z S_2^z$.
This transformation has no effect
on the expectation values of $z$-components of the spins but
leads to oscillatory behavior
of the $x$ or $y$ components, reflecting the fact
that the spins are rotating about the $z$-axis.
In the following it is implicitly assumed
that the basis states of the spins refer to
states in the corresponding rotating frame, even if we use the same
notation for the basis states.

NMR uses radiofrequency electromagnetic pulses to rotate the
spins.\cite{SLICHTER,BAYM}
By tuning the frequency
of the RF-field to the precession frequency of a particular
spin, the power (= intensity times duration) of the applied pulse
controls how much the spin will rotate. The axis of the
rotation is determined by the direction of the applied RF-field
(see Refs.~\citen{SLICHTER,BAYM}).

\section{Grover's database search algorithm}

Finding a particular entry in an unsorted list of $N$ elements
is a basic problem of searching databases. In general this takes
of the order of $N$ operations on a conventional computer.
It has been shown that a QC can find the item using
only \ORDER{\sqrt{N}} attempts.\cite{GROVERzero,GROVERone}

Consider the extremely simple case of a database containing
four items and functions $f_i(x)$ that upon query of the database
return minus one for the particular item we are searching for
and plus one otherwise.
Assuming a uniform probability distribution for the item
to be in one of the four locations, the average number
of queries required by a conventional algorithm is 9/4.
With Grover's QA the correct answer can be found
in a single query (this result only holds for
a database with 4 items).
Grover's algorithm for the four-item database can be
implemented on a 2-qubit QC.

The key ingredient of Grover's algorithm is an
operation called ``inversion about the mean'' that replaces each amplitude
of the basis states in the superposition by
two times the average amplitude minus
the amplitude itself. This allows then for the amplification
of the amplitude of the basis state that represents the
searched-for item. To see how this works it is useful to consider an example.
Let us assume that the item to search for corresponds
to e.g. number 2 ($f_2(0)=f_2(1)=f_2(3)=1$ and $f_2(2)=-1$).
Using the binary representation of integers with the order of the bits reversed,
the QC is in the state (up to an irrelevant phase factor as usual)
\begin{equation} \label{eq:psi}
\KET{\Psi}={1\over 2}(\KET{\uparrow\uparrow}+\KET{\downarrow\uparrow}-\KET{\uparrow\downarrow}+\KET{\downarrow\downarrow})
.
\end{equation}
We return to the question of how to prepare this state below.
The operator $D$ that inverts states like~(\ref{eq:psi}) about their mean reads
\begin{equation} \label{dmat}
D={1\over2}\pmatrix{
-1&\phantom{-}1&\phantom{-}1&\phantom{-}1 \cr
\phantom{-}1&-1&\phantom{-}1&\phantom{-}1 \cr
\phantom{-}1&\phantom{-}1&-1&\phantom{-}1 \cr
\phantom{-}1&\phantom{-}1&\phantom{-}1&-1 \cr }
\quad;\quad\matrix{\KET{\uparrow\uparrow} \cr \KET{\downarrow\uparrow} \cr \KET{\uparrow\downarrow} \cr \KET{\downarrow\downarrow} \cr}
.
\end{equation}
The mean amplitude (i.e. the sum of all amplitudes divided
by the number of amplitudes)
of~(\ref{eq:psi}) is 1/4 and we find that

\begin{equation}
D\KET{\Psi}=\KET{\uparrow\downarrow},
\end{equation}
i.e. the correct answer, and
\begin{eqnarray}
D^2\KET{\Psi}&=&
{1\over 2}(\KET{\uparrow\uparrow}+\KET{\downarrow\uparrow}+\KET{\uparrow\downarrow}+\KET{\downarrow\downarrow}), \\
D^3\KET{\Psi}&=&
-{1\over 2}(\KET{\uparrow\uparrow}+\KET{\downarrow\uparrow}-\KET{\uparrow\downarrow}+\KET{\downarrow\downarrow})=-\KET{\Psi},
\end{eqnarray}
showing that (in the case of 2 qubits)
the correct answer (i.e. the absolute value of the amplitude of $\KET{\downarrow\uparrow}$ equal
to one) is obtained after 1, 4, 7, ... iterations.
In general, for more than two qubits,
more than one application of $D$ is required to get the correct answer.
In this sense the 2-qubit case is somewhat special.

The next task is to express the preparation and query steps
in terms of elementary rotations. For illustrative purposes
we stick to the example used above.
We assume that initially the QC is in the state
with both spins up ($\KET{\uparrow\uparrow}$).
\footnote{We follow the convention used earlier in this paper,
i.e. the one used in Ref.~\citen{CHUANGtwo}) and
therefore deviate from the notation used in Ref.~\citen{CHUANGthree}}).
Transforming $\KET{\uparrow\uparrow}$ to the linear superposition~(\ref{eq:psi}) is a two-step process.
First the QC is put into the uniform superposition state:
\begin{equation}
\KET{U}= W_2W_1\KET{\uparrow\uparrow}
=-{1\over 2}(\KET{\uparrow\uparrow}+\KET{\downarrow\uparrow}+\KET{\uparrow\downarrow}+\KET{\downarrow\downarrow})
,
\end{equation}
where
\begin{equation}
W_j=X_jX_j{\bar Y}_j=
-{\bar X}_j {\bar X}_j{\bar Y}_j=
{i\over\sqrt{2}}\pmatrix{1&\phantom{-}1\cr1&-1\cr}_j
,
\end{equation}
\begin{equation}
X_j \equiv e^{i\pi S^x_j /2 \hbar} = \frac{1}{\sqrt{2}}
                       \left( \begin{array}{rr} 1 & i \\ i & 1 \end{array} \right)_j
\end{equation}
and
\begin{equation}
{\bar Y}_j \equiv e^{-i\pi S^y_j /2 \hbar} = \frac{1}{\sqrt{2}}
                       \left( \begin{array}{rr} 1 & -1 \\ 1 & 1 \end{array} \right)_j
\end{equation}
represents the Walsh-Hadamard (WH) transform
on qubit $j$ which transforms
$\KET{\uparrow}$ to $i(\KET{\uparrow}+\KET{\downarrow})/\sqrt{2}$,
a {\bf clock-wise} rotation of spin $j$ about $\pi/2$ around the $x$-axis and
a {\bf anti clock-wise} rotation of spin $j$ about $\pi/2$ around the $y$-axis respectively.
The inverse of a rotation $Z$ is denoted by $\bar Z$.
The second step is to encode the information in the database,
e.g. $f_2(x)$, in the state of the QC. This can be accomplished by
a transformation $F_i$ that in the case of our example $f_2(x)$ takes
the form
\begin{equation}
F_2=\pmatrix{
\phantom{-}1&\phantom{-}0&\phantom{-}0&\phantom{-}0\cr
\phantom{-}0&\phantom{-}1&\phantom{-}0&\phantom{-}0\cr
\phantom{-}0&\phantom{-}0&-1&\phantom{-}0\cr
\phantom{-}0&\phantom{-}0&\phantom{-}0&\phantom{-}1\cr}
\quad;\quad\matrix{\KET{\uparrow\uparrow}\cr\KET{\downarrow\uparrow}\cr\KET{\uparrow\downarrow}\cr\KET{\downarrow\downarrow}\cr}
.
\end{equation}
This transformation can be implemented by first letting the
system evolve in time:
\begin{eqnarray} \label{eq:zz}
I(\pi)\KET{U}&=&
e^{-i\pi S_1^z S_2^z}
\left[
{1\over 2}(\KET{\uparrow\uparrow}+\KET{\downarrow\uparrow}+\KET{\uparrow\downarrow}+\KET{\downarrow\downarrow})
\right] \nonumber \\
&=&
{1\over 2}(e^{-i\pi/4}\KET{\uparrow\uparrow}+e^{+i\pi/4}\KET{\downarrow\uparrow}
\nonumber \\
&&+e^{+i\pi/4}\KET{\uparrow\downarrow}+e^{-i\pi/4}\KET{\downarrow\downarrow}).
\end{eqnarray}
For the NMR-QC based on hamiltonian~(\ref{eq:ham}) this means
letting the system evolve in time (without applying pulses)
for a time $\tau_0 =-\pi/J_{1,2,z}$ (recall $J_{1,2,z}<0$).

Next we apply a sequence of single-spin rotations
to change the four phase factors such that we get the desired
state. The two sequences
$Y X {\bar Y}$ and $Y{\bar X}{\bar Y}$
are particulary useful for this purpose.
We find
\begin{eqnarray} \label{eq:yx}
\lefteqn{Y_1 X_1 {\bar Y}_1 Y_2{\bar X}_2{\bar Y}_2
\left[
{1\over 2}(e^{-i\pi/4}\KET{\uparrow\uparrow}+e^{+i\pi/4}\KET{\downarrow\uparrow}
\right.} \nonumber \\
&&\left.+e^{+i\pi/4}\KET{\uparrow\downarrow}+e^{-i\pi/4}\KET{\downarrow\downarrow})
\right]\cr
&=&
{1\over 2}(e^{-i\pi/4}\KET{\uparrow\uparrow}+e^{-i\pi/4}\KET{\downarrow\uparrow}
 \nonumber \\ &&
+e^{+3i\pi/4}\KET{\uparrow\downarrow}+e^{-i\pi/4}\KET{\downarrow\downarrow})\cr
&=&
{e^{-i\pi/4}\over 2}(\KET{\uparrow\uparrow}+\KET{\downarrow\uparrow}
-\KET{\uparrow\downarrow}+\KET{\downarrow\downarrow}).
\end{eqnarray}
Combining~(\ref{eq:zz}) and~(\ref{eq:yx}) we can construct the sequence
transforming the uniform superposition
in the state that corresponds to $f_i(x)$:
\begin{subeqnarray}
F_0&=&Y_1 {\bar X_1} {\bar Y}_1 Y_2 {\bar X_2}{\bar Y}_2 I(\pi), \\
F_1&=&Y_1 {\bar X_1} {\bar Y}_1 Y_2 {     X_2}{\bar Y}_2 I(\pi), \\
F_2&=&Y_1 {     X_1} {\bar Y}_1 Y_2 {\bar X_2}{\bar Y}_2 I(\pi), \\
F_3&=&Y_1 {     X_1} {\bar Y}_1 Y_2 {     X_2}{\bar Y}_2 I(\pi).
\end{subeqnarray}
The remaining task is to express the process of inversion about the mean,
i.e. the matrix $D$ (see~(\ref{dmat})), by a sequence of
elementary operations. It is not difficult to see that
$D$ can be written as the product of a
WH transform, a conditional phase shift $P$ and another WH transform:
\begin{eqnarray}
D&=&W_1 W_2 P W_1 W_2\cr
&=&W_1 W_2
\pmatrix{
\phantom{-}1&\phantom{-}0&\phantom{-}0&\phantom{-}0\cr
\phantom{-}0&-1&\phantom{-}0&\phantom{-}0\cr
\phantom{-}0&\phantom{-}0&-1&\phantom{-}0\cr
\phantom{-}0&\phantom{-}0&\phantom{-}0&-1\cr}
W_1 W_2.
\end{eqnarray}
The same approach that was used to implement $f_2(x)$ also works
for the conditional phase shift $P$ ($=-F_0$)
and yields
\begin{equation}
P=
Y_1{\bar X}_1{\bar Y}_1 Y_2 {\bar X}_2{\bar Y}_2 I(\pi).
\end{equation}
The complete sequence $U_i$ reads
\begin{equation}
U_i=W_1 W_2 P W_1 W_2 F_i
.
\end{equation}
Each sequence $U_i$ can be shortened by
observing that in some cases a rotation is followed by its inverse.
As there are various representations
of the WH transform $W_i$ that accomplish the same task, the sequence for e.g. $i=2$
can be written as
\begin{eqnarray}
W_1 W_2 F_2&=&
-{\bar X}_1 {\bar X}_1{\bar Y}_1 {    X}_2 {     X}_2{\bar Y}_2
Y_1 {X_1} {\bar Y}_1 Y_2{\bar X}_2{\bar Y}_2 I(\pi) \nonumber \\
&=&
-{\bar X}_1 {\bar Y}_1 { X}_2 {\bar Y}_2 I(\pi)    .
\end{eqnarray}
The sequences for the other cases can be shortened as well,
yielding
\begin{subeqnarray} \label{eq:udef}
&U_0={     X}_1{\bar Y}_1 {     X}_2{\bar Y}_2 I(\pi)
{     X}_1 {\bar Y}_1 {     X}_2 {\bar Y}_2 I(\pi), & \\
&U_1={     X}_1{\bar Y}_1 {     X}_2{\bar Y}_2 I(\pi)
{     X}_1 {\bar Y}_1 {\bar X}_2 {\bar Y}_2 I(\pi), & \\
&U_2={     X}_1{\bar Y}_1 {     X}_2{\bar Y}_2 I(\pi)
{\bar X}_1 {\bar Y}_1 {     X}_2 {\bar Y}_2 I(\pi), & \\
&U_3={     X}_1{\bar Y}_1 {     X}_2{\bar Y}_2 I(\pi)
{\bar X}_1 {\bar Y}_1 {\bar X}_2 {\bar Y}_2 I(\pi), &
\end{subeqnarray}
where in $U_1$ and $U_2$ we have dropped a physically irrelevant
sign.
Note that the binary representation of $i$ translates into the presence $(1)$
or absence $(0)$ in~(\ref{eq:udef}) of a bar on the rightmost $X_1$ and $X_2$.
\begin{table}[h]
\caption{Specification of the EO's
of a mathematically perfect 2-qubit QC.
The execution time of each
micro instruction is given by the second row ($\tau/2\pi$).
The inverse of e.g. ${\bar X}_1$ is found by reversing
the sign of $h_{1,x,0}$.
Model parameters omitted are zero for all EO's.} \label{tab:1}
\begin{tabular}{@{\extracolsep{\fill}}|c|ccccc|}
\noalign{\hrule}
\rule[-4pt]{0pt}{12pt}&$X_1$&${\bar X}_2$ & $Y_1$ & ${\bar Y}_2$ & $I(\pi)$ \\
\noalign{\hrule}
\rule[-4pt]{0pt}{12pt}
$\tau/2\pi$ & $0.25$ & $0.25$ & $0.25$ & $0.25$ & $50\times 10^{4}$ \\
\noalign{\hrule}
\rule[-4pt]{0pt}{12pt}
$J_{1,2,z}$ & $0$ & $0$ & $0$ & $0$ & $-10^{-6}$ \\
\noalign{\hrule}
\rule[-4pt]{0pt}{12pt}
$h_{1,x,0}$ & $+1$ & $0$ & $0$ & $0$ & $0$ \\
\rule[-4pt]{0pt}{12pt}
$h_{2,x,0}$ & $0$ & $-1$ & $0$ & $0$ & $0$ \\
\rule[-4pt]{0pt}{12pt}
$h_{1,y,0}$ & $0$ & $0$ & $+1$ & $0$ & $0$ \\
\rule[-4pt]{0pt}{12pt}
$h_{2,y,0}$ & $0$ & $0$ & $0$ & $-1$ & $0$ \\
\noalign{\hrule}
\end{tabular}
\end{table}
\begin{fulltable}[t!]
\caption{Specification of the elementary operations
implementing a 2-qubit NMR QC, using the notation of Table 1.
Note that a RF-pulse along the $y$($x$) direction corresponds to
a rotation about the $x$($y$).
Model parameters omitted are zero for all EO's.} \label{tab:2}
\begin{fulltabular}{@{\extracolsep{\fill}}|c|ccccc|}
\noalign{\hrule}
\rule[-4pt]{0pt}{12pt}
&$X_1$ & ${\bar X}_2$ & $Y_1$ & ${\bar Y}_2$ & $I(\pi)$ \\
\noalign{\hrule}
\rule[-4pt]{0pt}{12pt}
$\tau/2\pi$ & $10$ & $40$ & $10$ & $40$ & $50\times 10^{4}$ \\
\noalign{\hrule}
\rule[-4pt]{0pt}{12pt}
$J_{1,2,z}$ & $-10^{-6}$  & $-10^{-6}$  & $-10^{-6}$ & $-10^{-6}$ & $-10^{-6}$\\
\noalign{\hrule}
\rule[-4pt]{0pt}{12pt}
$h_{1,z,0}$ & $1$ & $1$ & $1$ & $1$ & $1$ \\
\rule[-4pt]{0pt}{12pt}
\rule[-4pt]{0pt}{12pt}
$h_{2,z,0}$ & $0.25$ & $0.25$ & $0.25$ & $0.25$ & $0.25$ \\
\noalign{\hrule}
\rule[-4pt]{0pt}{12pt}
$h_{1,x,1}$ & $0$ & $0$ & $0.05$ & $-0.05$ & $0$ \\
\rule[-4pt]{0pt}{12pt}
$h_{2,x,1}$ & $0$ & $0$ & $0.0125$ & $-0.0125$ & $0$ \\
\rule[-4pt]{0pt}{12pt}
$f_{1,x}$ & $0$ & $0$ & $1$ & $0.25$ & $0$ \\
\rule[-4pt]{0pt}{12pt}
$f_{2,x}$ & $0$ & $0$ & $1$ & $0.25$ & $0$ \\
\noalign{\hrule}
\rule[-4pt]{0pt}{12pt}
$h_{1,y,1}$ & $-0.05$ & $0.05$ & $0$ & $0$ & $0$ \\
\rule[-4pt]{0pt}{12pt}
$h_{2,y,1}$ & $-0.0125$ & $0.0125$ & $0$ & $0$ & $0$ \\
\rule[-4pt]{0pt}{12pt}
$f_{1,y}$ & $1$ & $0.25$ & $0$ & $0$ & $0$ \\
\rule[-4pt]{0pt}{12pt}
$f_{2,y}$ & $1$ & $0.25$ & $0$ & $0$ & $0$ \\
\noalign{\hrule}
\end{fulltabular}
\end{fulltable}
\begin{table}[h]
\caption{Final state of the qubits after running
the Grover's database search algorithm on an
ideal QC ($Q_1,Q_2$, model parameters given in table~\ref{tab:1}),
on a NMR-QC ($\hat Q_1,\hat Q_2$, model parameters given in table~\ref{tab:2})
and on the same NMR-QC ($\tilde Q_1,\tilde Q_2$, model parameters given in
table~\ref{tab:2})
using a different, but logically equivalent,
initialization sequence.}  \label{tab:3}
\begin{tabular}{@{\extracolsep{\fill}}|c|cccc|}
\noalign{\hrule}
\rule[-4pt]{0pt}{12pt}
& $U_0\KET{U}$&$U_1\KET{U}$ &$U_2\KET{U}$ & $U_3\KET{U}$  \\
\noalign{\hrule}
\rule[-4pt]{0pt}{12pt}
${Q_1}$ & $0.000$ & $1.000$ & $0.000$ & $1.000$  \\
\rule[-4pt]{0pt}{12pt}
${Q_2}$ & $0.000$ & $0.000$ & $1.000$ & $1.000$  \\
\noalign{\hrule}
\rule[-4pt]{0pt}{12pt}
${\hat Q_1}$ & $0.028$ & $0.966$ & $0.037$ & $0.955$  \\
\rule[-4pt]{0pt}{12pt}
${\hat Q_2}$ & $0.163$ & $0.171$ & $0.836$ & $0.830$  \\
\noalign{\hrule}
\rule[-4pt]{0pt}{12pt}
${\tilde Q_1}$ & $0.955$ & $0.041$ & $0.971$ & $0.027$  \\
\rule[-4pt]{0pt}{12pt}
${\tilde Q_2}$ & $0.031$ & $0.026$ & $0.971$ & $0.972$  \\
\noalign{\hrule}
\end{tabular}
\end{table}

\section{Results}
We now consider two different implementations of the Grover's QA
described above. The first and most obvious one is to make use of
theoretically ideal unitary transformations to perform the EO's.
One of the many possible choices for the model parameters that
correspond to this case are given in Table~\ref{tab:1}.
The second will be physical, i.e.
we will use the simulator to carry out the NMR-QC experiment itself.
The list of model parameters can be found in Table~\ref{tab:2}.
The values of the qubits $(Q_1,Q_2)$ are given by

\begin{equation}
Q_i\equiv {1\over2}-\langle S^z_i\rangle
,
\end{equation}

The numerical values of the qubits in the final state
as obtained by running Grover's QA on the simulator
are summarized in Table~\ref{tab:3}.
>From the data in the first two rows it is evident that
this QA performs as expected when the ideal EO's are used.
In the ideal case, the final state $(Q_1,Q_2)$
is the binary representation of the integer index of the ``-1'' item.

The third and fourth row contain the data for the NMR-QC case.
Using RF-pulses instead of ideal transformations
to perform $\pi/2$ rotations leads to less certain answers: The final
state is no longer a pure basis state but some linear superposition of the four
basis states.
Indeed, using a time-dependent external pulse to rotate spins
only yields an approximation to the simple rotations envisaged in theoretical work.
This affects the expectation values of the spin operators.
What is beyond doubt though is that it is easy to read off
the correct answer from the expectation values of the qubits.
Clearly the simulator reproduces the experimental results and
the QA seems to return the correct answer.

\begin{figure}
\epsfxsize=8cm\epsfysize=6cm\epsfbox{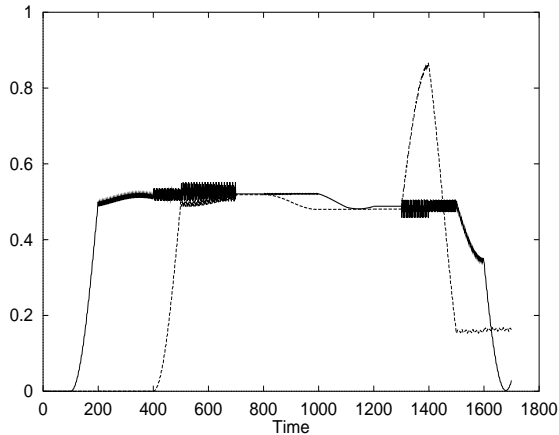}
\caption{Time evolution of the qubits $Q_1$ (solid line) and
$Q_2$ (dashed line) obtained by executing $W_1$, $W_2$ and sequence (4.18a)
for the case of the NMR-QC.
In all figures the time intervals for each operation have been
rescaled to make them look equal.}
\label{fig:1}
\end{figure}

\begin{figure}
\epsfxsize=8cm\epsfysize=6cm\epsfbox{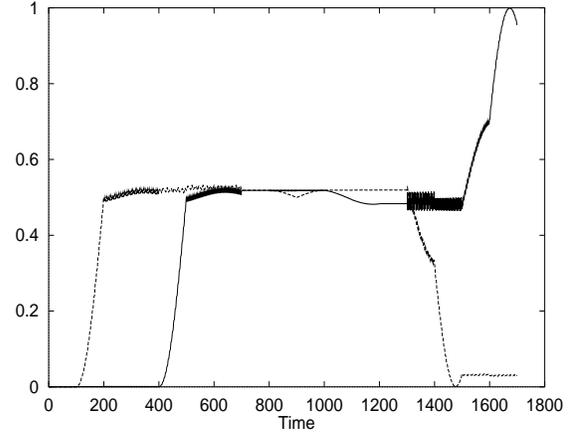}
\caption{Time evolution of the qubits $\tilde Q_1$ (solid line) and
$\tilde Q_2$ (dashed line) obtained by executing $W_2$, $W_1$ and sequence (4.18a)
for the case of the NMR-QC. Interchanging
the order in which the single-qubit operations $W_1$ and $W_2$
are applied changes the final state of the QC.}
\label{fig:2}
\end{figure}

\begin{figure}
\epsfxsize=8cm\epsfysize=6cm\epsfbox{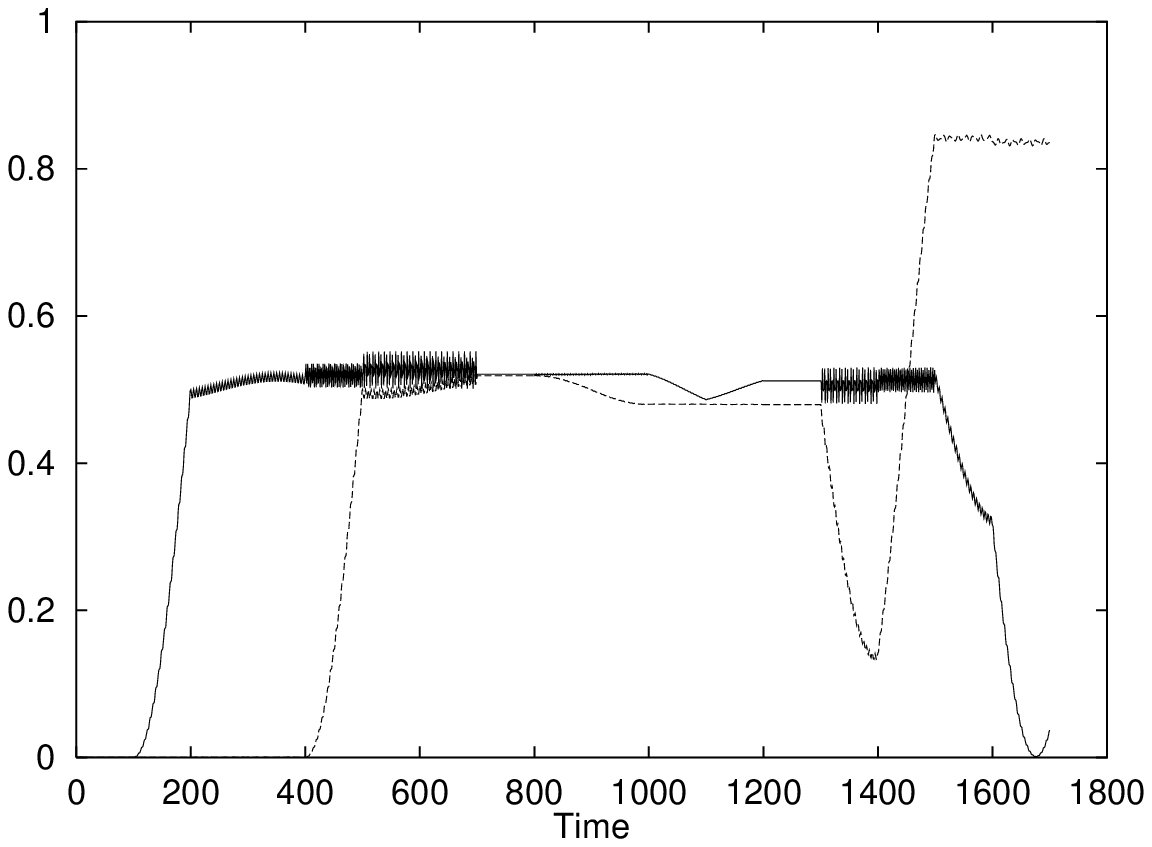}
\caption{
Same as Fig.1 but instead of sequence (4.18a) we used (4.18c).}
\label{fig:3}
\end{figure}

\begin{figure}
\epsfxsize=7.5cm\epsfysize=6cm\epsfbox{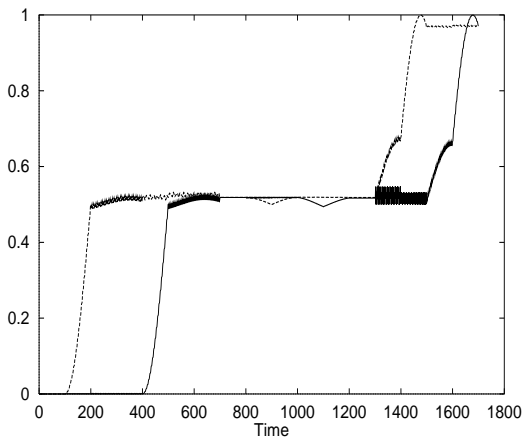}
\caption{
Same as Fig.2 but instead of sequence (4.18a) we used (4.18c).}
\label{fig:4}
\end{figure}

In an NMR experiment, application of each RF-pulse affects all spins in the sample.
Although the response of a spin to the RF-field
will only be large when this spin is at resonance, the state of the spins that are not
in resonance will also change. These successive unitary transformations
not necessarily commute with each other. If they do not commute
the state after these transformations depends on the order in which
the pulses have been applied.
In fact the presence of these perturbations, although small, can have
a devastating effect on the stability of the computation.
This is illustrated by the data in the fifth and sixth row
of Table~\ref{tab:3} and by Figs.~\ref{fig:1}-~\ref{fig:4}.
These results have been obtained by changing
the order of preparing the two spin states.
Instead of $W_1W_2$ we used $W_2W_1$ to initialize the QC,
a permutation that has no effect in the case of ideal EO's.
>From Table~\ref{tab:3} it is clear that in the case of EO's implementing
the 2-qubit NMR QC making this interchange leads to complete
wrong results. Many of such examples can be constructed: The very fact
that we cannot isolate one spin from the rest and perform
operations on the former only leads to phase errors that
may (but sometimes don't) alter the outcome of the calculation completely.
This QC architecture is intrinsically unstable to minor
modifications of the QA that are allowed from logical point of view,
a much more severe problem than that of ``decoherence''.
We believe it might be interesting to investigate these instabilities
experimentally.

Can the phase errors discussed above be (partially) eliminated by some clever
error-correction scheme?
At present there is no indication they can: Any error-correction method
requires adding extra spins to the system. The phase shift incured by the individual
spins will contribute to the phase shifts of each of the
many-body basis states and unless some magic cancelation
takes place, the final result is unlikely to be more stable.
On the other hand these unwanted phase factors are the result
of using RF pulses that only approximately implement rotations
about 90 degrees and may be reduced by using pulses that are more
complicated than the sinusoidal ones.
Perhaps dissipation effects may also help to reduce the sensitivity
to phase errors, a possibility that we are currently investigating.
Another route to more stable operation might be to use a different
set of EO's that more closely implements the ideal transformations.
For instance, non-adiabatic transitions between two levels driven
by a periodic field display peculiar
behavior\cite{SEIJIone} and might be employed to manipulate
the two-level systems.
In this respect the single-Cooper-pair-box\cite{NAKAMURAone}
may hold some promise. In this solid-state device
a non-adiabatic transition mechanism is used to let
a Cooper pair tunnel between two states.
Obviously there are many physical mechanisms to control the dynamics
of quantum spin systems.
Exploring which of these mechanisms is useful for quantum computing
may be a fertile area for future research.

\section*{Acknowledgements}
Generous support from the Dutch ``Stichting Nationale Computer
Faciliteiten (NCF)''
and from the Grant-in-Aid for Research of the
Japanese Ministry of Education, Science and Culture
is gratefully acknowledged.

\end{document}